\documentclass[prl,twocolumn]{revtex4}
\usepackage{graphicx}

\usepackage{ifthen}
\usepackage{xcolor}
\usepackage{mathrsfs}
\usepackage{bm}
\usepackage{yfonts}
\usepackage{chapterbib}
\usepackage{amsmath}
\usepackage{amssymb}
\usepackage{amsfonts}
\usepackage{upgreek}
\usepackage{wasysym}
\usepackage{mathbbol}

\usepackage{ulem}

\definecolor{DarkBlue}{rgb}{0.0,0,0.6}
\definecolor{DarkRed}{rgb}{0.6,0,0.0}
 
\DeclareMathAlphabet{\mathpzc}{OT1}{pzc}{m}{it}

\newcommand{\be}{\begin{equation}}
\newcommand{\ee}{\end{equation}}
\newcommand{\bea}{\begin{eqnarray}}
\newcommand{\eea}{\end{eqnarray}}

\newcommand{\br}{{\bf r}}

\newcommand{\vexr}{v_\text{ex}({\bf r})}
\newcommand{\Vex}{\hat V_\text{ex}}
\newcommand{\HH}{\hat H}
\newcommand{\Nd}{N_\text{d}}
\newcommand{\eF}{\varepsilon_\text{F}}
\newcommand{\GR}{\Gamma_\mathcal{R}}
\newcommand{\GL}{\Gamma_\mathcal{L}}
\newcommand{\GRks}{\Gamma^\text{KS}_\mathcal{R}}
\newcommand{\GLks}{\Gamma^\text{KS}_\mathcal{L}}
\newcommand{\SL}{\Sigma_\mathcal{L}}
\newcommand{\SR}{\Sigma_\mathcal{R}}
\newcommand{\ed}{\varepsilon_\text{d}}

\graphicspath{{./Figures/}}

\begin{document}


\title{DFT-based transport calculations, Friedel's sum rule and the
  Kondo effect}
\author{Philipp Tr{\"o}ster{{$^1$}}}
\author{Peter Schmitteckert{$^{2,3}$}}%
\author{Ferdinand Evers{$^{2,3,4}$}}%
\affiliation{{$^1$ Institut f\"ur BioMolekulare Optik, 
   Ludwig-Maximilians-Universita\"at M\"unchen,
   D-80538 Munich, Germany}}%
\affiliation{$^2$ Institute of Nanotechnology, Karlsruhe Institute of
   Technology, D-76021 Eggenstein-Leopoldshafen, Germany}%
\affiliation{$^3$Center of Functional Nanostructures, 
   Karlsruhe Institute of Technology, D-76131 Karlsruhe, Germany}%
\affiliation{$^4$
   Institut f\"ur Theorie der Kondenserten Materie, Karlsruhe
   Institute of Technology, D--76128 Karlsruhe, Germany}%
 
\date{\today}

\begin{abstract} 
Friedel's sum rule  provides an explicit expression for a conductance functional,
$\mathcal{G}[n]$, valid for the single impurity Anderson model at zero
temperature.  The functional is special because it 
does not depend on the interaction strength $U$. As a consequence, the 
Landauer conductance for the Kohn-Sham (KS) particles of density
functional theory (DFT) coincides with the true
conductance of the interacting system. The argument breaks down at
temperatures above the Kondo scale, near integer filling,
$n_{\text{d}\sigma}\approx 1/2$ for spins
$\sigma{=}\uparrow\downarrow$. 
Here, the true conductance is strongly suppressed by the Coulomb
blockade, while the KS-conductance still indicates resonant
transport. Conclusions of our analysis are corroborated by 
DFT studies with numerically exact exchange-correlation functionals 
reconstructed from calculations employing the density matrix
renormalization group. 
\end{abstract}


\maketitle %
The ground state density functional theory
(DFT) owes its success to the fact that
it proves enormously useful in the prediction of electronic
properties of molecules, solids and surfaces.
\cite{dreizler90,fiolhais03}
Therefore, applications towards the electronic
transport properties of single molecules and self
assembled monolayers came together quite naturally with the 
corresponding experimental successes in the field.
\cite{koentopp08,scheerBook} 
Besides being useful for quantitative calculations,
more fundamental properties of DFTs and the corresponding
exact functionals have been an issue of intensive research. 
As an example we mention the question, what exactly is the nature
of the approximations when using ground state DFT in combination
with the Landauer formalism for transport calculations.
\cite{stefanucci:195318,koentopp:121403}

The Landauer approach formulated in terms of 
non-equilibrium Green's functions\cite{brandbyge02,diVentraBook}
and its validity for Kohn-Sham particles is also 
our topic in this work. 
It relates a single particle Hamiltonian, $H^\text{KS}$, 
to the conductance via the
transmission function, $G=T^\text{KS}(\eF) e^2/h$. 
The definition is 
\be
\label{e1}
T^\text{KS}(E) = \text{Tr}\  \GRks(E) G(E) \GLks(E) G^\dagger(E)
\ee
where 
$\Gamma^\text{KS}_\alpha=\mathfrak{i}(\Sigma_\alpha-\Sigma^\dagger_\alpha)$, 
$G=(E-H^\text{KS}-\SR-\SL)^{-1}$ 
and the trace is over the Hilbert space associated with $H^\text{KS}$. 
The self energies $\Sigma_\alpha$ describe the coupling of the KS-system 
to external reservoirs, $\alpha=\mathcal{L,R}$. 
They are given by a golden-rule expression,
$\Sigma_\alpha(E)=|V|^2g^\text{KS}_\alpha(E)$,  where $V$ is the
coupling matrix element and $g^\text{KS}_\alpha$ is a Green's function of
the  leads. \cite{haug96} 

Often quantum transport is dominated by a single orbital of the molecule or
the quantum dot (QD), only. Therefore, in studies of correlated 
electron transport  interacting
level models are standard, 
e.g., the single impurity Anderson model (SIAM, see also (\ref{e5}))
\be
\label{e2}
 \HH_\text{QD} {=} \epsilon_\text{d} \hat \Nd+ 
    U \left( \hat{n}_{\mathrm{d}\uparrow} - \frac{1}{2} \right) \left(
      \hat{n}_{\mathrm{d}\downarrow} 
- \frac{1}{2} \right), 
\ee
where $\hat \Nd = \hat n_{\mathrm{d},\uparrow}+\hat n_{\mathrm{d},\downarrow}$
with $\hat n_{\mathrm{d},\sigma} =
\hat{d}_{\sigma}^{\dagger}
\hat{d}_{\sigma}^{\phantom{g}}$
and spin $\sigma=\uparrow,\downarrow$. 
In such models an analogue version of (\ref{e1}) is valid 
featuring retarded and advanced Green's functions 
of the interacting system ($\Gamma_{\alpha,\sigma\sigma'}=\Gamma_\alpha\delta_{\sigma\sigma'}$),
\be
\label{e3}
T(E) = \frac{\GL\GR}{\GL+\GR} \mathcal{A}_{\text{d}}(E)
\ee 
where we have introduced the spectral function of the interacting QD,
$
\mathcal{A}_{\text{d}}(E) {=} \mathfrak{i}\text{Tr}_\sigma
\left(G^\text{r}(E){-}G^\text{a}(E)\right)$. 
\cite{meir92}

By comparing  Eq. (\ref{e3}) and (\ref{e1}) one might suspect, 
that in order to accurately reproduce the
true value  for the transmission, $T(\eF)=T^\text{KS}(\eF)$, 
it is neccessary for the KS-theory to also reproduce the true spectral
function, $\mathcal{A}_\text{d}(E)$. 
It is easy to see, that the latter is not possible,
however, unless $U{=}0$. 
To this end we recall that
$\mathcal{A}_\text{d}(E)$,  
carries the two Hubbard peaks at energies
{$\epsilon_\text{d} \pm U/2$}.\cite{bruusBook}
These peaks are not seen  by the KS-system 
because the model (\ref{e2}) does not exhibit magnetism,
so both spin channels, ($\uparrow,\downarrow$), are equivalent.
Therefore, 
 $H^\text{KS}$ is diagonal, 
$H^\text{KS}_\text{d}=\ed^\text{KS} \delta_{\sigma\sigma'}$,
and the KS-spectral function, 
$\mathcal{A}^\text{KS}=\mathfrak{i}\text{Tr}_\sigma(G-G^\dagger)$,
supports a single peak centered about 
$\ed^\text{KS}$, only. 
Despite of the absense of the Hubbard peaks in $\mathcal{A}^\text{KS}$
we argue that in addition to the ground
state density also the KS-conductance coincides
with the true value of the interacting system, $T=T^\text{KS}$.

Moreover, we maintain that this statement is correct even though
$H^\text{KS}$ is not unique in the sense that the
exchange-correlation on-site potential, 
$v^\text{KS}_\text{d}=\ed^\text{KS}-\ed$, 
can be complemented by an XC-contribution to the 
couplings, $V\rightarrow V^\text{KS}$, 
as well. Different combinations ($V^\text{KS},v^\text{KS}_x$) 
produce an effective single particle
Hamiltonian with the correct ground state density. 
In fact, as will be demonstrated below, 
the on-site exchange-correlation (XC) 
potential and the coupling to the leads can be
drastically different. Correspondingly, the resonance position,
$\ed^\text{K}$, and broadenings, $\Gamma_\mathcal{L,R}$,
that determine  $\mathcal{A}^\text{KS}$ will be strongly XC-functional 
dependent. 
We will see that despite of this ambiguity in $H^\text{KS}$ 
the KS-conductance is an observable that  takes a unique value coinciding with 
the true conductance. 


We start with a general recollection about features of 
ground state DFT that follow directly from
the first Hohenberg-Kohn theorem. \cite{hohenberg64}
According to this theorem, we can reconstruct the
external potential, $\vexr$
that an interacting gas of $N$ electrons
is exposed to, if the ground state density
$n(\br)$ together with the Hamiltonian
$\hat H_0$ in the absense of
any $\vexr$ is known.
Thus we reproduce the full Hamiltonian
$\HH=\hat H_0 + \Vex$ from our knowlege of $n(\br)$
(up to a constant shift in energy). 

Knowing $\HH$ we can calculate, in principle,
all equilibrium response
functions of the $N$-particle system, 
provided that the ground state is unique, once
$n(\br)$ has been specified.
Hence, we can consider such correlators
to be a functional of the ground state density. 
\footnote{The functionals are  uniquely defined, 
once the kinetic part of $\HH^\text{KS}$, i.e. the piece that remains
invariant under changes in  $v_\text{ex}(\br)$ and $n(\br)$, is fixed.}
For example the dynamical
density-density susceptibility, $\chi[n](\br,\br^\prime,t-t')$,
can be thought of being such functional of $n(\br)$. 
Correlators can also be calculated
  at finite temperature. Again,
  they can be thought of as a unique functional of the
  ground state density under the condition that 
  the equilibrium state does not exhibit
  a broken symmetry which is not already
  encoded in $n(\br)$. 

A correlation function of special interest to us is the
current-current correlation function, i.e. the conductivity.
Its longitudinal part describing the response to potential gradients
is closely related to the density susceptibility:
\be
-\mathfrak{i}\omega \chi[n](\br,\br',\omega) =
\frac{\partial}{\partial \br}
\frac{\partial}{\partial \br^\prime}
\sigma_\ell[n](\br,\br^\prime,\omega)
\ee
The susceptibility, in turn, may be expressed as a conductance
in transport experiments that operate with electrodes for which the
Fermi-liquid description holds true. Hence, in such geometries
also the conductance is a functional of the ground state density:
${\mathcal G}[n]$. Hence,  this conductance can be calculated using the
ground state DFT if the proper functional ${\mathcal G}[n]$ is being
used.

The previous statement is as correct as it is useless for practical
purposes unless a
good approximation for ${\mathcal G}[n]$ can be given. Of course, even
if such an approximation would be known, in practice calculations
would still suffer from inaccuracies in approximate XC-functionals 
used to obtain $n(\br)$. 

{\bf Friedel's sum rule for the SIAM.} 
The complete definition of the SIAM-Hamilonian reads \cite{hewsonBook}
\begin{eqnarray}
\label{e5}
\HH &=& \HH_\text{QD} +\sum_{\alpha=\mathcal{L,R}} \HH_{\alpha} +
\HH_\text{T}  \\
  \HH_{\alpha} &{=}& {-}t \sum_{x=1, \sigma}^{M  {-} 1}
  \left( \hat{c}_{x+1,\sigma,\alpha}^{\dagger}
  \hat{c}_{x,\sigma,\alpha}^{\phantom{g}} {+} \text{h.c.}\right),\\
  \HH_\text{T} &{=}& - V\sum_{\sigma,\alpha} \left( \hat{c}_{1,\sigma,\alpha}^{\dagger} \hat{d}_{\sigma}^{\phantom{g}} + \hat{d}_{\sigma}^{\dagger} \hat{c}_{1,\sigma,\alpha}^{\phantom{g}} \right) \mathrm{.} 
\end{eqnarray}
The $\hat{c}_{x,\sigma,\alpha}^{(\dagger)}$ denote fermionic annihilation
(creation) operators at site $x$, lead $\alpha=\mathcal{R,L}$ 
The model is of interest to us because
it affords the Friedel sum rule\cite{hewsonBook}, 
\be
\label{e8}
\mathcal{A}_{\text{d}\sigma}(\eF) = \frac{\sin^2[\pi n_{\text{d}\sigma}(\eF)]}{\Gamma/2}.
\ee
where $\Gamma=\GL+\GR$. By combination with (\ref{e3})
we relate the particle number 
on the quantum dot, $\Nd/2=n_{\text{d}\uparrow}=n_{\text{d}\downarrow}$,  to the conductance,  
\be
\label{e9}
\mathcal{G}[n] = \frac{2 e^2}{h}  
\frac{\GL\GR}{(\Gamma/2)^2}\  \sin^2\left( \frac{\pi}{2} \Nd \right), 
\ee
The identity constitutes an exact analytical expression
for a conductance functional ${\mathcal G}[n]$.%
It is remarkable that in the case of symmetric coupling, 
$\GL{=}\GR$, Eq.  (\ref{e9}) relates the conductance associated with 
 $\HH$ to a single 
system characteristics, only, 
which is the particle number on the QD, $\Nd. $
This implies, that any change in the parameters of $\HH$ leaves the 
conductance invariant, provided that $\Nd$ is unchanged 
and that the conditions of
applicability of (\ref{e9}) are still valid. 
One requirement for this is that both leads are 
(effectively) non-interacting and free of backscattering,
so that the Fermi-liquid picture holds a zero temperature.   
The last condition is imporant because else 
$\pi n_{\text{d}\sigma}$ would not properly account for the
number of bound states (per spin) and the scattering phase.

{\em Friedel's sum rule and ground state DFT.} 
The KS-Hamiltonian of ground state DFT reads 
\be
\HH^\text{KS} =\epsilon^\text{KS}_\text{d} \hat \Nd +
\sum_{\alpha} \left[ \HH_\alpha +  \sum_{x=1, \sigma}^{M} v^\text{KS}_{x,\alpha} \hat N_{x,\alpha} \right]
+ \HH_\text{T} 
\ee
where the onsite XC-potentials,  
$\epsilon^\text{KS}_d[N]=\epsilon_\text{d}+v^\text{KS}_\text{d}[N]$ 
and $v^\text{KS}_{x,\alpha}[N]$, are functionals of the 
local particle density $N_x$. 
At first sight, the KS-Hamiltonian, $\hat \HH^\text{KS}$, does not 
meet the requirements for the validity of  $(\ref{e9})$, 
since the XC-functional extends into the leads and thus might
contribute to the scattering phase shift: 
$\delta(\eF)\to \delta^\text{KS}(\eF)$.
However, this impression is misleading as can be seen from the
following argument. The Friedel sum rule in its general form relates 
the extra scattering phase shift, $\delta(\eF)$, 
induced by decreasing $\ed$ (down from $\infty$) to 
the extra spectral weight, $\Delta\mathcal{A}(E)$, 
thus generated: 
\be
\label{e11}
\frac{\delta(\eF)}{\pi} = \int^{\eF}_{-\infty} \frac{dE}{2\pi} \
\Delta \mathcal{A}_\sigma(E) = \Delta N_\sigma
\ee
The right hand side denotes the total change in the
particle number per spin, 
$\Delta N_\sigma=\sum_{x} \Delta N_{x\sigma}$, 
associated with the occupation 
of this extra weight.  A special aspect of Eq. (\ref{e8}) 
is the implication that the sum rule is
exhausted already by the density change on the QD, 
$\Delta N_\sigma=N_\text{d}/2$: the net charge 
accumulated by 
bringing $\ed$ down is entirely concentrated 
in the QD. 
This latter statement must be correct also for the 
KS-theory, by definition. Since also KS-particles obey 
the relation (\ref{e11}) we conclude: 
$\delta(\eF)=\delta^\text{KS}(\eF)=\pi N_\text{d}/2$ and
Eq. (\ref{e9}) holds for them as well despite of the presence of 
$v^\text{KS}_{x\alpha}$ in the leads. 

\begin{figure}[thb]
	\centering
		\includegraphics[width=0.48\textwidth]{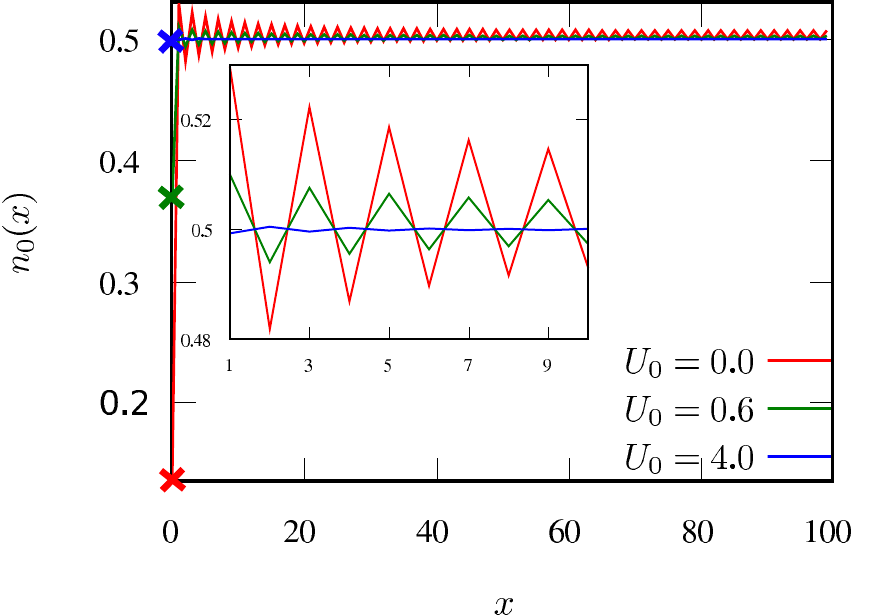}
		\caption{Ground state density per spin of a QD (at $x=0$) coupled
                  to a non-interacting reservoir with $M{=}100$ sites for growing
                  on-QD interaction  $U = 0.0, 0.6, 4.0$. Parameters: 
                  $\ed{=}0.2, V=0.3$, bandwidth of conduction
                  electrons: $2t=2$. }
		\label{f1}
\end{figure}
Functional (\ref{e9}) is valid 
at any value of the interaction strength $U$ and at
temperatures below the Kondo scale, 
$T_\text{K}\sim 2t \sqrt{2\Gamma/U}e^{-U/4\Gamma}$
near integer filling $N_\text{d}\approx 1$.%
\cite{hewsonBook} In this context 
the Abrikosov-Suhl resonance underlying the Kondo-effect 
plays a crucial role. At temperatures above $T_\text{K}$ it is not 
developed and the conductance is strongly suppressed due to the 
Coulomb blockade; relation (\ref{e9}) is strongly violated. 
The XC-functional of DFT must be very sensitive to 
Kondo-physics. This is obvious for the following reason: 
we have seen that KS-transport reproduces the exact transmission. 
To reproduce the resonant transport (Kondo) scenario in the 
regime $N_\text{d}\approx 1$, 
the KS-level of the quantum dot must be half filled for each spin, 
implying $|\eF-\ed^\text{KS}|\apprle \Gamma$ in the Kondo regime
even if the bare position of the level $-\ed \gg \Gamma$.

To illustrate and extend our analysis we have calculated the 
ground state density and corresponding exact XC-functionals employing 
the density matrix renormalization group method and 
``backward'' DFT. The approach has proven useful before in the
context of the interacting resonant level model (IRLM). \cite{schmitteckert08} 
We have adapted  our technology here to treat the SIAM. \cite{troester10}. 

In our calculations we consider a coupling to a single lead only in
order to reduce the computational effort. As long as the ground state
is concerned the case with two leads and 
symmetric couplings, $\GL=\GR$, 
has an exact mapping into the single lead case, 
essentially because the odd combination of tunneling operators, 
$c^{(\dagger)}_{\text{odd},1}=(
c^{(\dagger)}_{\mathcal{L},1}-c^{(\dagger)}_{\mathcal{R},1})/\sqrt{2}$
decouples from the QD.

\begin{figure}[thb]
	\centering
		\includegraphics[width=0.48\textwidth]{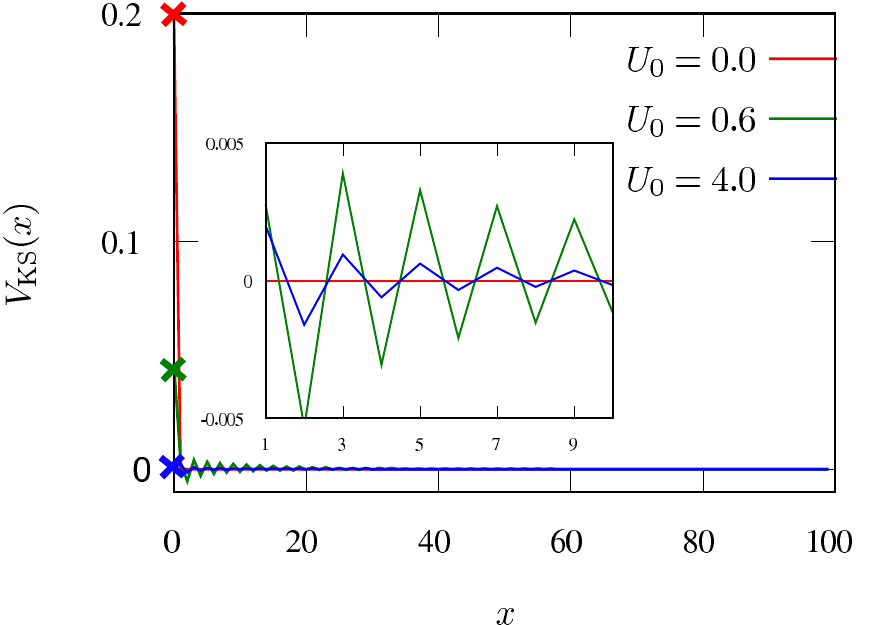}
		\caption{XC-correlation potential corresponding to the
                  evolution of the density shown in Fig. \ref{f1}. 
                  The on-dot potential is denoted
                  $v^\text{KS}_0=\ed^\text{KS}$. 
}
		\label{f2}
\end{figure}
In Fig. \ref{f1} we display the evolution of the density in a
QD with $\ed$ slightly above $\eF$. Without interactions, the QD is
empty. Upon increasing $U$, the dot fills because in
the spirit of a Jellium model we have defined the interaction in 
{\eqref{e2}} with respect to density fluctuations against a 
background $n^\text{bg}_{\text{d}\sigma}=1/2$. 
The density in the leads, $x\geq 1$, exhibits the typical 
Friedel-oscillations with their $2k_\text{F}$ periodicity and the 
$1/x$-envelop. Their amplitude is controlled by the boundary
condition which is set by the QD. Since its occupation changes
from $N_\text{d}\approx 0$ to $N_\text{d}\approx 1$  
the phase shift associated with backscattering 
increases by $ 2 \pi n_{\text{d},\sigma}$. This is why at large $U$
the Friedel-oscillations are anti-phase with the case $U{=}0$, 
inset Fig. (\ref{f1}). We witness a signature of
Kondo-physics, cf. \cite{affleck08}.

Fig. \ref{f2} shows the KS-potential on the
QD, $\ed^\text{KS}$and in the lead, $v^\text{KS}_{x}$, 
corresponding to the evolution of the density, Fig. \ref{f1}.  
On the dot $v^\text{KS}_{x}$  shows the expected behavior from 
                 $\ed^\text{KS}=\ed$ at $U=0$ to 
                 $\ed^\text{KS}\approx 0$ at large interaction, here
                 $U=4$. 
In the leads the potential oscillations follow the
Friedel-oscillations of the density and introduce the interaction 
corrections. The oscillation amplitude depends in a non-monotonic
way on $U$, increasing from the non-interacting fixpoint and
decreasing again when approaching the strong-coupling, Kondo
fixpoint, see inset. 

In a KS-theory based on a local KS-potential, the hybridzation $V$
coincides with the one from the original coupling $\hat H_\text{T}$. 
\begin{figure}[thb]
	\centering
		\includegraphics[width=0.3\textwidth]{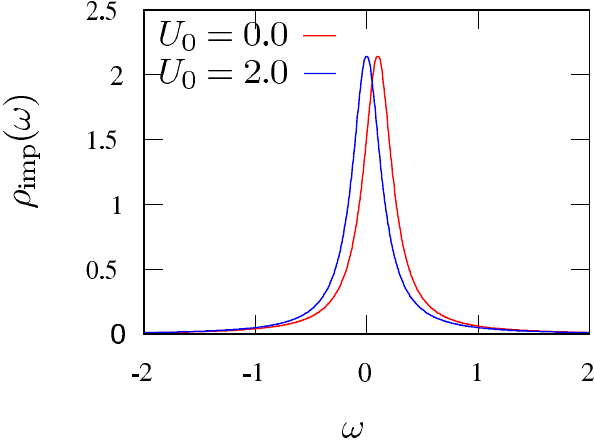}
		\caption{Local spectral function of the KS-system, 
                  $\mathcal{A}^\text{KS}_{d}=2\pi\rho_\text{imp}$,
                  on the QD for interactions $U=0,2$. 
                 Parameters: $\ed=0.1$, $V=0.15...$, $\eF=0$.  }
		\label{f3}
\end{figure} 
This suggests that the on-dot spectral function,
$\mathcal{A}^\text{KS}(E)$  of the KS-system  has
a width  close to the non-interacting one, $\Gamma\approx \Gamma^\text{KS}$. 
Fig. \ref{f3} fully supports this point of view. In addition, it also shows
that indeed there is only a single maximum (Hubbard peaks don't exist
in KS-theory)  which will be getting closer to $\eF$ with increasing interaction
strength $U$. 

\begin{figure}[htp]
	\centering
		\includegraphics[width=0.48\textwidth]{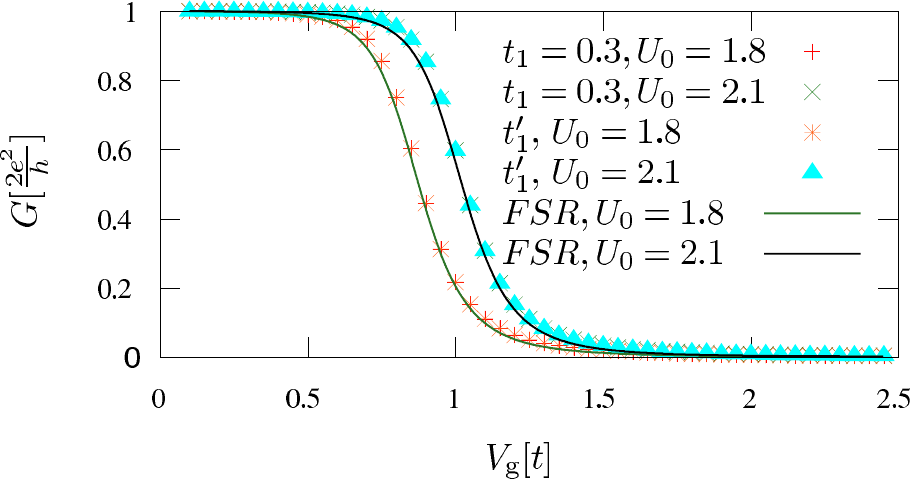}
		\caption{Comparison of conductances calculated via the
                  Friedel sum rule (FSR)  and directly from the 
                  Landauer formula for KS-particles, (\ref{e12}). 
                  Parameters $V=0.3$, $U=1,8,2.1$ are considered. }
		\label{f4}
\end{figure}

Fig. \ref{f4} compares exact transmission curves obtained via the
Friedel sum rule Landauer's formula ($\GL=\GR$)
\be
\label{e12} 
T(\eF) = \frac{(\Gamma/2)^2}{(\ed^\text{KS}-\eF)^2+(\Gamma/2)^2}
\ee 
The good agreement between the results obtained with both methods
illustrates the point emphasized above:
Even though the KS-spectral function is not physical,
the associated transmission is close to the precise value. 

We briefly mention that this point can be highlighted further
by constructing a version of
KS-theory, in which not only the diagonal elements of the density matrix, 
i.e. $n_{x\sigma}$  but also off-diagonal elements are faithfully 
reproduced. This can be achieved by 
adding to the on-site potential  also  a
modification  of hopping matrix elements, i.e. $V \to {V^\text{KS}}^\prime$
{such that the expectation value of the kinetic energy of the impurity coupled to
the first lead site within the KS description matches the one obtained from DMRG}; 
for the technical details see \cite{troester10}.  
Responding to this change  the width of 
the KS-spectral function of the modified theory
is no longer close to the original one, 
$\Gamma^\text{KS} \to {\Gamma^\text{KS}}^\prime$. 
Since the charge in the QD must remain unchanged,  we expect 
a compensating shift in the on-site energy, $\ed^\text{KS} \to {\ed^\text{KS}}^\prime$. As can be seen from
Fig. \ref{f4} despite a substantial change, $V=0.3$ is replaced by
$V=0.1...$, the transmission when evaluated via (\ref{e12}) is not
changed.  In view of Eq. (\ref{e12}) this finding is easily
understood:  $T(\eF)$ is determined by the same ratio,
$\Gamma^\text{KS}/(\ed^\text{KS}-\eF)$, that also fixes the density. 
Since by construction in all KS-models the density is the same, 
we have 
$\Gamma^\text{KS}/(\ed^\text{KS}-\eF) =
{\Gamma^\text{KS}}^\prime/({\ed^\text{KS}}^\prime-\eF)$, 
so the transmission remains the same, also.  

{\bf Conclusions.} 
We briefly discuss two generalizations of the preceeding analysis. 
First, our argument was assuming a single level, only, while 
a real QD, e.g. a molecule, exhibits in general several levels. 
One expects, however, that the main conclusion remains valid as long
as all levels contribute independently to the transport current. In
particular,  the sum rule (\ref{e9}) should remain a useful
approximation for the true density functional. 
Second, our analysis heavily relies on the Kondo effect restoring full
transmission in the case of single occupation of the dot. 
Suppose, that the Kondo temperature is very low, and that the
measurement is done at slightly higher temperatures. Then transport
is dominated by the Coulomb blockade and the conductance is strongly 
suppressed,  $T(\eF)\sim (\Gamma/U)^2\lll 1$. On the other hand 
the particle density is essentially still the ground state one,
$n(\br)$, i.e. it is largely  insensitive to this change and in particular
$n_{\text{d}\sigma}\approx 1/2$.  We conclude that in this case the
Friedel sum rule (\ref{e9}) does not hold and that KS-theory
(without breaking spin rotational invariance)  does not
reflect this change, i.e. we still have: $T^\text{KS}(\eF)\approx 1$. 

\acknowledgments
We acknowledge support by the Center of Functional Nanostructures at
KIT.  We also thank Gianlucca Stefanucci and Peter W\"olfle for
discussions. After completing our work we have learned about an
independent research by Bergfield, Liu, Burke and Stafford
\cite{bergfield11}.
Where they overlap, their conclusions coincide with the ones
presented in this and an earlier publication \cite{ESsubmitted} 
\bibliographystyle{apsrev}
\bibliography{../BibME/bibDft,../BibME/bibDftApplTransport,../BibME/bibInelastic,../BibME/bibOwnMolEl,../BibME/bibNegf,../BibME/bibTDFT,../BibME/books,../BibME/bibGeneral,../BibME/bibMethods,./p25.Friedel}

\end{document}